# Investigation of superconducting properties of NbN films deposited by DC magnetron sputtering on high-k dielectric HfO$_2$ buffer layer

Porokhov N.V.[1], Sirotina A.P.[1], Pershina E.A.[1], Shibalov M.V.[1], Diudbin G.D.[1], Mumlyakov A.M.[1], Timofeeva E.R.[1], Trofimov I.V.[1], Tagachenkov A.M.[1], Anufriev Yu.V.[1], Zenova E.V.[1] and Tarkhov M.A.[1]

[1]Institute of Nanotechnology of Microelectronics of the Russian Academy of Sciences, Nagatinskaya 16a-11, Moscow 115487, Russia

E-mail: nporokhov@gmail.com

## Abstract

The influence of the buffer layer of hafnia dielectric (HfO$_2$) on superconducting properties of niobium nitride films (NbN), produced by the technique of the reactive magnetron depositing has been investigated for the first time. This study presents a comprehensive analysis of morphological, microstructural, and electrophysical parameters of thin NbN films. The dependence of transition critical temperature from the thickness of the HfO$_2$ buffer layer has been obtained. For the first time, it has been demonstrated that the superconducting properties of niobium nitride films at the HfO$_2$ buffer layer have high superconductive parameters, namely, T$_c$ ~ 13 K (at the film thickness of approximately 30 nm) and high critical current density of approximately 10$^7$ A/cm$^2$. The relation between the thickness of the HfO$_2$ buffer layer with the critical current density of niobium nitride films has been determined. It appeared that the values of critical current density increase significantly up to 13 MA/cm$^2$ when the thickness of the hafnia buffer layer is more than 2 nm.

## 1. Introduction

Niobium nitride is widely adopted in the manufacturing of superconducting nanoelectronic devices of various functionalities, such as single-photon detectors SNSPD[1][2], hot-electron bolometers (HEB) and mixers of THz frequency range[3], microwave kinetic inductance detectors (MKIDs)[4], etc. Thin NbN films are produced using various techniques, such as reactive magnetron sputtering with Nb target in Ar and N$_2$[5] gas mixture, pulsed laser deposition (PLD)[6], high-temperature chemical vapor deposition (HTCVD), and atomic layer deposition (ALD)[7][8]. Generally, to ensure high operating parameters, all mentioned techniques require intensive heating of a substrate up to 600°C and more.

Critical current density ($J_c$) and critical superconducting transition temperature ($T_c$) are the integral parameters of a superconducting film. Microscopic parameters of a film, such as film thickness, size of particles[9], phase composition[10], and stoichiometry[11][12], the occurrence of micro- and macro-stress[13][14][15], number and type of defects, as well as the oxygen concentration in the NbN film, play an essential role in $T_c$ and $J_c$ values. Due to the correlation, defining the effect of each parameter on the superconductive properties of a film is rather complicated. It is reliably known that when decreasing the NbN film thickness, the superconducting transition temperature is also decreasing owing to the critical thickness of a film where the superconductive properties disappear[16][17]. Moreover, when decreasing the thickness of a polycrystalline film, not only the average size of a particle decreases, but the value of internal stress increases as well.

It's worth remarking that some applied issues involving ultrathin NbN films require depositing to amorphous materials $SiO_2$ or $Si_3N_4$, using the CVD technique. It is also known that depositing of NbN films to amorphous substrates and substrates with a considerable discrepancy in parameters of unit cells with NbN results in the deterioration of superconductive properties[6][18][19]. Thus, the production of ultrathin films (~ 10 nm) with high superconductive parameters requires the implementation of buffer layers for improvement of adhesive properties and providing conditions for epitaxial growth of films. In their studies [20][21], investigated the influence of the MgO buffer monocrystalline layer on the superconductive properties of NbN films. Sufficient improvement of superconductive parameters due to the considerable influence of the buffer layer on the structural characteristics of a film has been demonstrated. However, the hygroscopic properties of magnesium oxide lead to the deterioration of properties of such films.

Hafnium dioxide ($HfO_2$) can be used as another material for the buffer layer. Hafnium oxide has proven productive material in manufacturing nano-devices. Herewith, deposition of hafnium oxide requires a high temperature of a substrate[22]. Besides, hafnium oxide has numerous advantages. Foremost these are the high values of dielectric constant $k = 20 - 25$ and a band gap ~ 5.68 eV[23]. Currently, there is not any information on the influence of the thickness of oxide hafnium buffer layer on the superconductive properties of NbN films in the literature. This is precisely why this study is concerned with the investigation of superconductive properties, surface morphology, as well as microstructural and phase composition of thin NbN films on the $HfO_2$ buffer layer. NbN films have been produced using magnetron sputtering with

Nb target in Ar and N2 gas mixture, the HfO$_2$ sublayer has been deposited by atomic layer deposition.

High superconductive parameters of ultrathin NbN films are achieved when depositing using magnetron technique to hot substrates with a temperature from 600°C and higher. In this study, we investigate the properties of NbN films deposited at a relatively low temperature of substrates, approximately 300°C, as such temperature allows integrating the produced films into the sensitive processes of microelectronics.

## 2. Experimental details

The samples under investigation comprised a multilayer composition Si(n-type)/SiO2/HfO2/NbN. Amorphous silicon oxide with a thickness of 500 nm was deposited to a single-crystal silicon wafer of n-type with a thickness of 4 inches. Then the HfO$_2$ dielectric layer was produced in-situ using the technique of atomic layer deposition. After that, the NbN layer was deposited ex-situ using the magnetron sputtering technique. The niobium nitride films were produced with varied thicknesses of the buffer sublayer of HfO$_2$ dielectric. A schematic representation of the investigated samples is given in Figure 1.

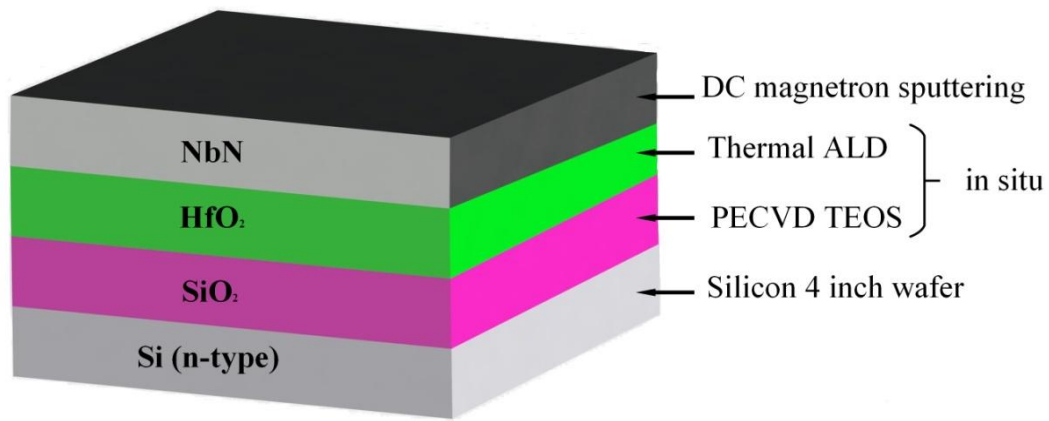

**Figure 1.** Schematic representation of the investigated samples.

The amorphous layer of silicon oxide was deposited using the PECVD technique from the TEOS silicon-containing precursor and oxygen. Before depositing, the wafer was processed in N$_2$O plasma for cleaning from organic contamination. The substrate temperature was 300°C, pressure in the chamber was 400 mTorr, argon flow through the TEOS was 50 sccm RF source power was 70 W. Hafnium oxide was deposited by atomic layer deposition. Hafnium oxide film was deposited by alternating reaction series including the feeding of the TDMAH (Hf(CH$_3$)$_2$N)$_4$) metalloorganic precursor and water as a reactant. The cycle duration of feeding the TDMAH and water was 400 ms and 10 ms, respectively. Before feeding the precursor into the chamber, feeder degasification was carried out to release excessive pressure in the feeder. High Purity Argon (99,9999%) acted as purging gas. The deposition was performed at the substrate temperature of 300°C. The temperature of the TDMAH precursor was 70°C to ensure the required pressure of saturated vapors. The niobium nitride film was deposited by reactive magnetron sputtering with niobium target (99,999% pure) in the inert gas mixture of argon and nitrogen. The ratio of argon and nitrogen was 4:1, respectively. The power of DC magnetron was 5 W/cm$^2$. The depositing speed under these conditions is 11 nm per minute. The substrate temperature during the deposition is 300°C.

For the purpose of investigation of the influence of the buffer layer thickness on the properties of niobium nitride films, a number of samples with various thicknesses of $HfO_2$ from 0 to 30 nm were produced. The thickness of a layer was determined using the ellipsometry technique[24]. The $SiO_2$ and NbN layers were applied under the same conditions from one sample to another. The layer thickness was 500 nm and 28 nm, respectively.

The first measured parameter, i.e., surface resistivity, illustrated the homogeneity of the NbN film on a 4-inch wafer, the value of the surface resistivity average parameter, and its spread at a room temperature. It has been established that the variation of resistivity along the wafer did not exceed 4%, which is due to the geometry of the confocal magnetron. When performing optimization of the relative orientation of a substrate, the variation of surface resistivity can be reduced up to 1% and less.

The surface morphology was investigated using an atomic-force microscope (AFM) operating in semi-contact and contact modes. The NbN surface roughness was analyzed on an area of 1 μm$^2$ in a semi-contact mode. After processing of results, parameters of roughness for defined areas were obtained, where $R_q$ – root mean square deviation of the surface height from the weighted average value, and $R_a$ – mean absolute deviation of the height value from the weighted average value. AFM images taken in a semi-contact mode for NbN films without a buffer layer (a) and with a buffer layer with a thickness $t_{HfO2}$ of 2 nm, 15 nm, and 30 nm (b, c, d) are given in Figure 2. It has been determined that $R_a$ and $R_q$ roughness parameters progressed almost 2 times with the increase of the thickness of the $HfO_2$ buffer layer.

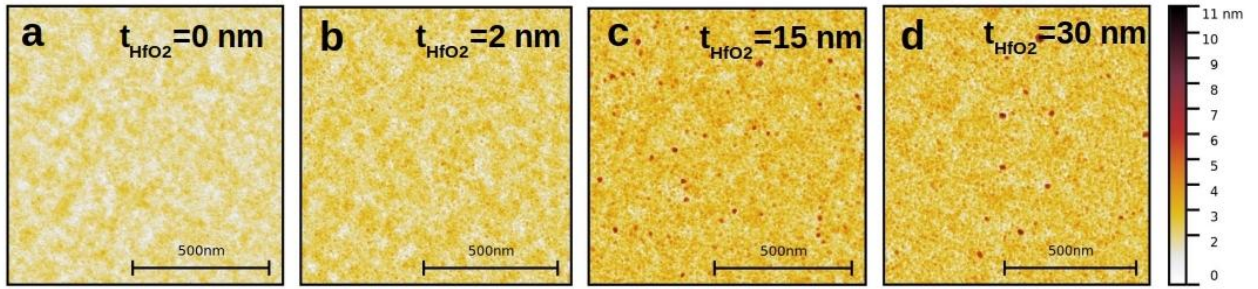

**Figure 2.** AFM images of the NbN films surfaces: (a) NbN on PECVD $SiO_2$, (b) NbN on PECVD $SiO_2$ / $HfO_2$ 2 nm thick, (c) NbN on PECVD $SiO_2$ / $HfO_2$ 15 nm thick, (d) NbN on PECVD $SiO_2$ / $HfO_2$ 30 nm thick.

According to the AFM images, the thickness of the $HfO_2$ buffer layer affects the morphology of the NbN film. A slight layer of hafnium oxide of 2 nm affects the morphology of the NbN film surface with a thickness of 28 nm, prompting the suggestion that a sublayer influences the character of the NbN film growth.

Critical current density and critical superconducting transition temperature are the integral parameters defining the quality of films produced. That is the reason why measuring the critical superconducting transition temperature is an intrinsic part of the characterization of NbN films [25]. The critical temperature was measured using the non-contact method of magnetic insulation of the obtained films. The technique is based on the Meissner-Ochsenfeld effect [26], the sample was placed between two coaxial induction coils. Alternating current was supplied to one of the coils. If the sample is under normal conditions alternating magnetic field will stimulate EMF from the induction coil in a signal coil. When transitting to a superconducting state, the sample shields the magnetic field, and the induced EMF disappears in the signal coil. Dependencies of the rated power in the signal coil from the temperature are given in Figure 3.

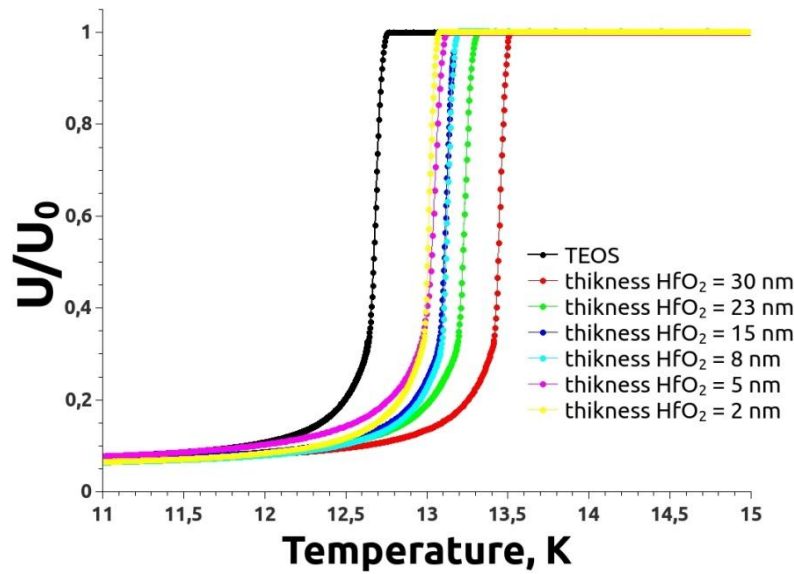

**Figure 3.** A curves family of the dependencies of the rated power in the signal coil from the temperature for a series of samples with different thicknesses of the buffer layer.

It is apparent from the charts that the thickness of the $HfO_2$ sublayer affects the critical transition temperature of the NbN film. The minimum temperature $T_c$ of the NbN film without the $HfO_2$ buffer layer was 12.65 K. When there is the $HfO_2$ buffer layer, the critical transition temperature of the NbN film increases monotonically up to 13.6 K. A precision temperature measuring diagram of the investigated sample was used in the critical temperature measuring diagram. It can be seen that an insignificant thickness of $HfO_2$ leads to a considerable temperature increase, while the subsequent increase of the buffer layer results in the monotonical increase of the superconducting transition temperature. According to the obtained results, the hafnium oxide buffer layer considerably affects the character of growth of the niobium nitride film even with a low thickness (~ 2-3 nm).

The dependency of superconducting transition temperature and the average value of $R_a$ roughness from the thickness of the hafnium oxide buffer layer is given in Figure 4. The dependency of critical temperature from the NbN film roughness is given in the insert. The obtained result is in good agreement with the results of the investigation of superconductive parameters typical of thick films [5]. It has been established that the rough surface is developed due to the increase of the size of film particles which indirectly implies the influence of the buffer layer on the particle size. It may also be assumed that increasing the thickness of the $HfO_2$ sublayer leads to the increase of the average size of $HfO_2$ crystallites, which in its turn increases the surface roughness degree. Eventually, subsequent conformal deposition of the NbN film to the surface with varying degrees of roughness results in monotonically increasing roughness of the NbN film surface and the increase of the thickness of the $HfO_2$ sublayer.

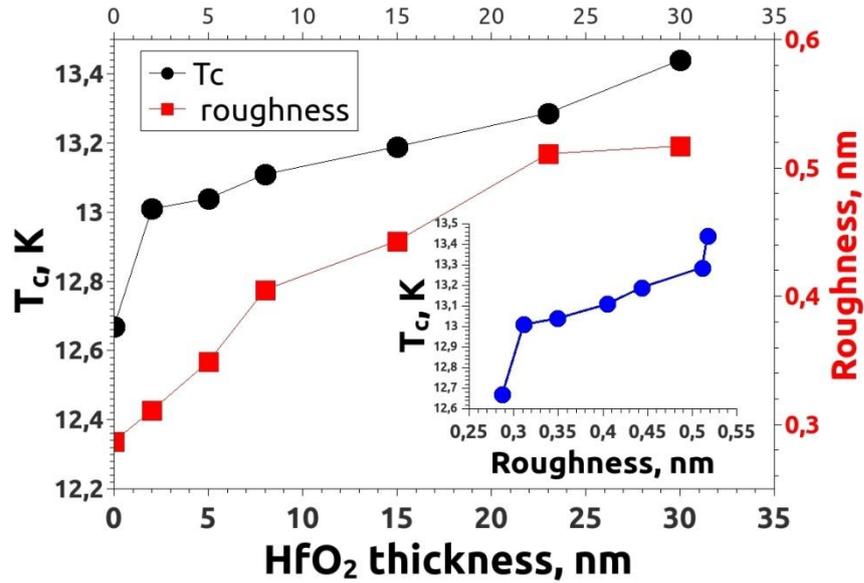

**Figure 4.** Dependences of superconducting transition temperature and the average value of $R_a$ roughness from the thickness of the hafnium oxide buffer layer. The dependency of critical temperature from the NbN film roughness is given in the insert.

For the purpose of investigation of the buffer layer influence on the critical current density, the micro-bridge structure with a width of 1 μm was made. The micro-bridge pattern was composed using the laser-based lithography method with subsequent plasma-chemical etching in the Ar/$SF_6$ gas mixture. Aluminum with a thickness of ~ 100 nm was used for the metallization of pads.

Dependency of current density ($J_c$) from the thickness of the hafnium oxide buffer layer at a temperature of 4.2K is given in Figure 5. As a temperature for investigating the dependency of the current density ($J_c$) from the thickness of the buffer layer the fixed point of 4.2K corresponding to the liquid helium boiling point was selected. The insert to Figure 5 contains the image of the investigated micro-bridges. Additional layout optimization of super-current input into the micro-bridge area has not been carried out.

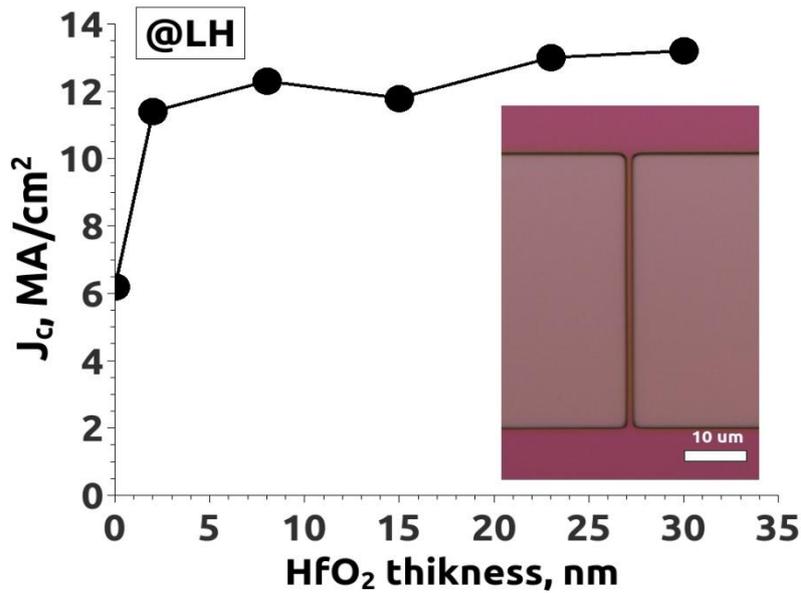

**Figure 5.** Dependency of current density ($J_c$) from the thickness of the hafnium oxide buffer layer at a temperature of 4.2K, insert contains image of the investigated micro-bridges.

The critical current density was defined by the current-voltage curve with reference to the micro-bridge topology. All investigated samples were cooled in the Gifford McMahon closed-cycle cryostat cooled to a temperature of 2.3K. Current-voltage curves were measured using the Keithley 6221A precision low noise current source operating in the current stabilization mode.

According to Figure 5, the presence of the buffer layer leads to a significant increase of the critical current density by a factor of two. However, the transient of the current density occurs at a low thickness of the buffer layer (~ 2 nm). The further increase of the buffer layer thickness results in an insignificant growth of the critical current density. These studies demonstrate the similarity of the behavior of the superconducting transition temperature and critical current density where there is the hafnium oxide buffer layer. This fact implies a substantial influence of the buffer layer on the NbN parameters.

Therefore, the investigation of microstructure and phase analysis have been performed. First, the influence of the sublayer thickness on the stoichiometry of the NbN films was investigated by the Auger electron spectroscopy using the ion beam etching. It has been determined that the film stoichiometry is the same both for the sample with the $HfO_2$ sublayer (30 nm) and without the $HfO_2$ sublayer. The precise stoichiometry of the NbN film was not analyzed in this study due to the complexity of the analysis of the thin NbN film composition. The development of the methodology for defining the precise stoichiometry of the thin NbN films will be the subject of the follow-up studies.

Next, the X-ray diffraction analysis of the produced films was performed. Grazing Incidence X-ray Diffraction (GIXRD) was obtained by X-ray diffractometer with the Bragg-

Brentano Focusing. The X-ray incident angle was 0.4°. The diffraction patterns were carried out in parallel geometry using CuK$_\alpha$ radiation, scanning was performed at angles 2θ. The scanning of the sample was through the range of 2θ angles 23-80°, scanning step was 0.03°. The coarse-grained aluminum film was selected as a standard sample for position calibration at an angle of 2θ (a=4.048±0.002Å).

The microstructure of the NbN film specimen has been investigated by transmission and high-resolution electron microscopy techniques (TEM and HRTEM) using the JEOL JEM-2100 Plus microscope at an accelerating voltage of 200 kV. The two-dimensional Fourier transform technique has been used to obtain diffraction patterns and to minimize the noise of high-resolution TEM images. The required sample preparation has been carried out for TEM investigation, i.e., lamellae in the cross-section of investigated films using the Focused Ion Beam technique (FIB). To enhance the image contrast and for convenience in future TEM investigations, a carbon layer was applied (about 30 nm). After that, to protect the specimen surface from the ion beam influence, a platinum layer was applied (about 150 nm). The layers mentioned above have not been considered in the structural analyses.

Grazing Incidence X-ray Diffractions for the number of samples of thin NbN films of various thicknesses on the HfO$_2$ sublayer are given in Figure 6(a). It can be concluded from Figure 6(a) that under such conditions of imaging there are no differences in the structure of samples. The lack of difference among the diffractions may be explained by the high FWHM value for peaks (≈1.5°) due to the small size of crystallites (≈5-7 nm) and possible internal stresses. A wide peak at ≈32° for samples with the HfO$_2$ sublayer thickness of 15 nm and more corresponds to the signal from HfO$_2$. According to the GIXRD data, the value of interplanar spacing for samples under investigation is 2.55; 2.22; 1.55-1.53; 1.33; 1.27 Å. The given set of interplanar spacings is mostly related both to δ-NbN phase (Fm-3m, PDF2 01-071-0300) and γ-Nb$_4$N$_3$ phase (I4/mmm, PDF2 01–089-6041). It is worth mentioning that when processing diffractions there is a deviation from experimental data, which is indicative both of distortions on the NbN lattice and by-product phase. Distortion of crystallite lattice in thin films of metal nitrides deposited at low temperatures is the result of a possible presence of a large number of vacancies (metal or nitrogen), as well as internal stresses [27]. Thus, the structure of thin NbN films can be represented as a distorted cubic lattice with the calculated parameter of lattice a = 4.40 Å.

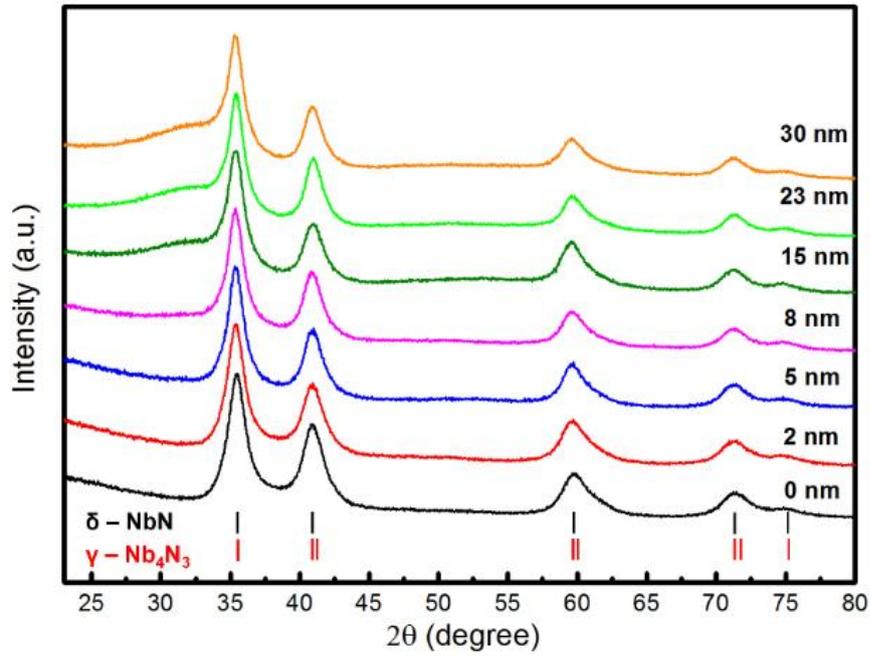

(a)

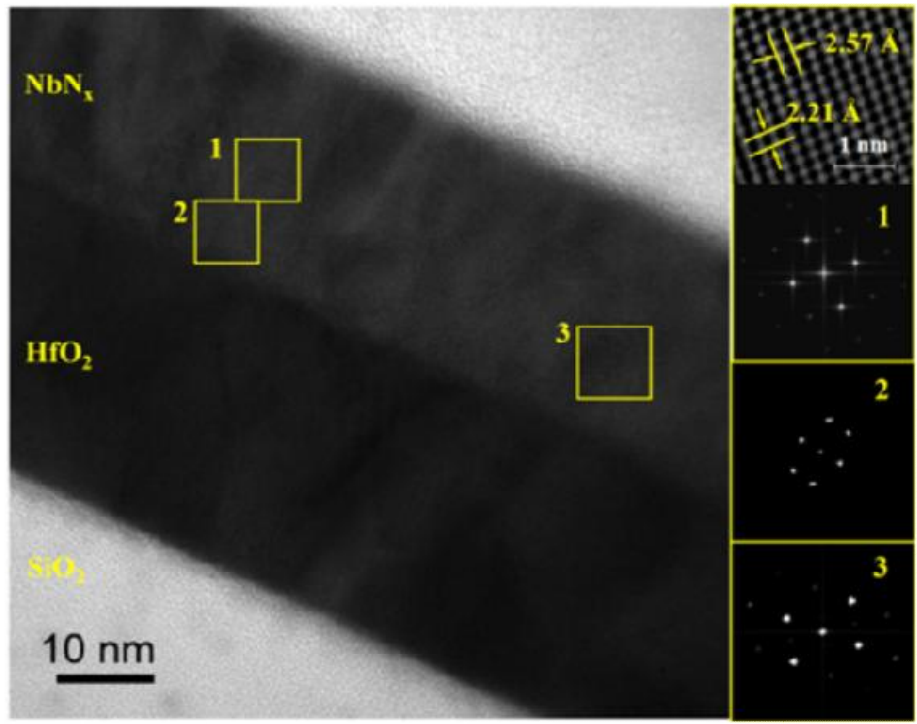

(b)

**Figure 6.** (a) Grazing incidence X-ray diffraction (GIXRD) vs. two theta scans from series NbN/HfO$_2$ (X nm)/SiO$_2$/Si(110) for different thickness of underlayer HfO$_2$. The grazing incident angle is fixed at 0.4° from the surface. Lines indicate position peak for δ-NbN (black lines) and γ-Nb$_4$N$_3$(red lines) b) TEM is an image of a cross section of an NbN film on an HfO$_2$ sublayer, HRTEM is an image of an NbN film. 1-3 insert corresponding FFT patterns obtained from the individual regions NbN film.

After that, to demonstrate the microstructure, the analyzed films were investigated by the TEM technique. For illustrative purposes, the structure of the NbN films produced by magnetron sputtering on the HfO$_2$ sublayer, on the surface of the SiO$_2$/Si is given in Figure 6(b). According to the Figure 6(b), the investigated films represent a sequence of several layers: amorphous SiO$_2$ (lower level), then a layer of deposited hafnium oxide, a layer of deposited niobium nitride, and a protective carbon layer.

The given film consists of conformal HfO$_2$ and NbN layers with thicknesses of 28 nm and 24 nm, respectively. According to the obtained HRTEM images, the average size of NbN crystals is 8 nm. The squares on the Figure 6(b) represent the areas of separate crystals, from which the Fourier transform for the analysis of phase composition was taken. The diffraction patterns from the marked areas of a film are given in Figure 6(b) on the right. The FFT images contain only separate bright reflections, there are not any wide diffuse peaks. This means that the NbN film under study is of a polycrystalline structure. Analysis of dark-field TEM images has demonstrated that some crystallites are of elongated shape in the direction perpendicular to the plane of the film surface. During the investigation of all films of the obtained series of samples, no differences in the microstructure of samples have been found. It should be noted that due to the polycrystalline structure of a sample and variation of crystallite sizes, as well as the non-homogeneous form of crystallites, the accuracy of determining the average size of a crystal is not more than 0.5 nm both according to XRD and HRTEM data. At the same time, the variation of average sizes of crystallites in 0.5 nm can explain the observable experimental dependence $T_c$ [16].

**Conclusion**

In this study, the influence of the buffer layer of hafnia dielectric (HfO$_2$) on superconducting properties of niobium nitride films (NbN) has been investigated. The hafnium oxide buffer layer considerably affects the character of growth of the niobium nitride film even with a slight thickness (~ 2-3 nm), increasing the critical superconducting transition temperature by 0.5K, and the critical current density by a factor of two. It appeared that the values of critical current density increase significantly up to 13 MA/cm$^2$ when the thickness of the hafnia buffer layer is more than 2 nm. Depending on the thickness of the HfO$_2$ buffer layer, structural characteristics of the NbN films did not vary significantly, which is due to the small size of the film crystallites.

**Acknowledgment**

The authors would like to thank the Doctor of Physical and Mathematical Sciences, A.M. Ionov and the Doctor of Physical and Mathematical Sciences V.V. Chernyshev for assistance in the investigation and meaningful remarks, E.M. Eganova for the preparation of samples for TEM investigation, V.S. Belov for assistance in investigating the samples by the Auger electron spectroscopy.

The study has been executed with the support of project No. 0004-2019-0004 of the Ministry of Education and Science of the Russian Federation. Fabrication and characterization were carried out at large scale facility complex for heterogeneous integration technologies and silicon+carbon nanotechnologies.